\pdfoutput=1
\documentclass[fleqn,usenatbib,usedcolumn]{mnras}
\usepackage[british]{babel}    
\usepackage{graphicx}
\usepackage{amsmath}
\usepackage{xspace}
\usepackage{ulem}
\usepackage{color}
\usepackage{dcolumn}

\newcommand\Eq[1]{(\ref{#1})}
\newcommand\Fig[1]{Fig.~\ref{#1}}
\newcommand\Tab[1]{Table~\ref{#1}}

\renewcommand{\to}{t_0}

\renewcommand{\Im}{\textrm{Im}}

\usepackage[T1]{fontenc}
\usepackage{ae,aecompl}

\usepackage{mathptmx}
\usepackage{txfonts}

\title[Fast quadrupole and hexadecapole computation]{Fast computation of quadrupole and hexadecapole approximations in microlensing with a single point-source evaluation}

\author[A. Cassan]{Arnaud Cassan\\
Institut d'Astrophysique de Paris, UPMC Univ Paris 6 et CNRS, Sorbonne Universit\'es, UMR 7095, 98 bis bd Arago, F-75014 Paris, France}

\date{Last updated <date>; in original form <date>}

\pubyear{2015}

\begin{document}
\label{firstpage}
\pagerange{\pageref{firstpage}--\pageref{lastpage}}
\maketitle

\begin{abstract}
The exoplanet detection rate from gravitational microlensing has grown significantly in recent years thanks to a great enhancement of resources and improved observational strategy. Current observatories include ground-based wide-field and/or robotic world-wide networks of telescopes, as well as space-based observatories such as satellites \textit{Spitzer} or \textit{Kepler/K2}. This results in a large quantity of data to be processed and analysed, which is a challenge for modelling codes because of the complexity of the parameter space to be explored, and the intensive computations required to evaluate the models. In this work, I present a method that allows to compute the quadrupole and hexadecapole approximation of the finite-source magnification with more efficiency than previously available codes, with routines about $\times\,6$ and $\times\,4$ faster respectively. The quadrupole takes just about twice the time of a point-source evaluation, which advocates for generalizing its use to large portions of the light curves. The corresponding routines are available as open-source \textsc{python} codes.
\end{abstract}




\section{Introduction}

	Since the visionary work of \cite{MaoPaczynski1991}, Galactic gravitational microlensing has led to the discovery of dozens of exoplanets and brown dwarfs,\footnote{\url{http://exoplanet.eu/catalog/}} and revealed an unexpected population of cold, low-mass exoplanets located beyond the snow-line of their stars. Statistical studies have more recently settled that exoplanets in the Milky Way are the rule rather than the exception \citep{Cassan2012}, thereby opening exceptional prospects to discover exoplanets in a variety of systems and configurations. A recent highlight of exoplanets microlensing search is the characterization of the mass and distance from Earth of planetary microlenses through space parallax measurements: such observations are performed simultaneously with ground-based observatories and from space, using \textit{Spitzer} \cite[e.g.][]{Udalski2015,Street2016} and \textit{Kepler/K2} \citep[campaign C9,  2016 April 7 through July 1,][]{Henderson2016}, with a strong involvement of the international microlensing community.
	
	The recent upgrades of ground-based telescopes, including robotic ones, have dramatically increased the amount of photometric data that need to be processed, with thousands of data points for which the models have to be computed. In fact,  modelling is currently the most difficult task in microlensing and in most cases the bottleneck of detections delivery. Hence, the improvement of both the strategy of the exploration of the parameter space and the efficiency of the computations are of prime interest, in their mathematical and numerical aspects.

	Improved strategies to search the parameter space first include the exploitation of features in the light curves to limit the region to be explored. \cite{Albrow1999} have introduced a model to fit individual caustic crossings independently from the whole light curve. This strategy has been extended to binary-lens caustic-crossing events through the definition of a specific parametrization (dates of caustic entry and exit and corresponding positions of the source centre on the caustics) that allows to limit the search to light curves producing caustic magnification peaks at the dates seen in the data \citep{Cassan2008,Cassan2010}. To avoid unnecessary calculations of light curves, \cite{Penny2014} has developed in a similar manner the concept of caustic regions of influence that are defined as empirical analytic expressions limiting the parameter space to regions where most low-mass planetary signals lie. A better understanding of the link between the caustic topography and the resulting light curves is a key ingredient to limit the region in the parameter space to explore. A detailed study of this aspect has been conducted by \cite{Liebig2015} in the binary lens case. In that work, the authors established a classification of all possible peaks in the light curves into four types only, and arranged the corresponding possible light-curve morphologies into 73 categories. Hence, inspecting the characteristics of the observed light curves (number and shape of peaks for example) naturally provides clever initial guesses for the subsequent minimization algorithms \citep[in particular Bayesian algorithms, e.g.][]{Kains2012}. 
	
	A second direction of improvement is to design more efficient mathematical methods and numerical codes to perform the calculations of the probed microlensing models. \cite{Skowron2012} have proposed a new algorithm to solve complex polynomial equations, which can solve the lens equation faster than classical roots finder codes. Nevertheless when finite-source calculations are required, the computation time significantly increases. This has triggered the development of a number of methods or approximations to overcome the problem. While magnification maps obtained by inverse ray-shooting are well suited for a large number of lenses  \citep{Wambsganss1997}, they are in general much too slow to be computed in real time, even for triple lenses. Pre-computed magnification maps are however useful for statistical studies where a large number of simulated light curves needs to be computed \citep[e.g.][]{Jovi2008,Cassan2012}. A refined image-centred ray-shooting algorithm has been developed by \cite{Bennett2010} to more specifically address the calculation of high-magnification planetary models, in which the source images are highly elongated. Image contouring provides an interesting alternative to ray-shooting \citep{GouldGaucherel1997,Dong2006,Dominik2007}, because it is far less demanding in computation time. \cite{Bozza2010} has significantly improved the contouring method by including an error control algorithm which optimizes the sampling of the contour of the images.

	When finite-source effects are noticeable but are weak enough (e.g. caustic or cusp passages without caustic crossing), multipole approximations of the finite-source magnification \citep{PejchaHeyrovsky2008} have proven to be of great help because the computation time is several order of magnitudes below that of the exact finite-source magnification. A simple implementation of the quadrupole and hexadecapole approximations has been proposed by \cite{Gould2008}, using respectively 9 and 13 (point source) resolutions of the polynomial lens equation to evaluate numerically the corresponding coefficients of the expansion.
	
	In this work, I present an improved implementation of the quadrupole and hexadecapole approximations, based on the construction of the image contours through a Taylor expansion around the individual images of the source centre, and which makes use of a single resolution of the lens equation. In section \ref{sec:mul}, I present the method that yields the multipole coefficients of the expansion, and in section \ref{sec:appli}, I discuss its implementation and numerical efficiency.

\section{Multipole expansion} \label{sec:mul}

\begin{table*}
\label{tab:Q}
\centering 
\begin{tabular}{p{0.7cm}p{4.5cm}|p{4.5cm}|p{4.5cm}|p{0.5cm}l}
\hline\\[-7pt]
& \centering$W_3$ & \centering$W_4$ & \centering$W_5$ & \centering$W_6$ & \\\\[-7pt]
\hline\\[-5pt]
$Q_{2,0}$ & $a_{1,0}^2$ &\centering---&\centering---&\centering---&\\\\[-5pt]
$Q_{1,1}$ & $a_{1,0}\;a_{0,1}$ &\centering---&\centering---&\centering---&\\\\[-5pt]
\hline\\[-5pt]
$Q_{3,0}$ & $3\;a_{1,0}\;a_{2,0}$ & $a_{1,0}^3$  &\centering---&\centering---&\\\\[-5pt]
$Q_{2,1}$ & $2\;a_{1,0}\;a_{1,1} + a_{0,1}\;a_{2,0}$ & $a_{1,0}^2\;a_{0,1}$  &\centering---&\centering---&\\\\[-5pt]
\hline\\[-5pt]
$Q_{4,0}$ & $4\;a_{1,0}\;a_{3,0} + 3\;a_{2,0}^2$ & $6\;a_{1,0}^2\;a_{2,0}$ & $a_{1,0}^4 $ &\centering---&\\\\[-5pt]
$Q_{3,1}$ & $3\;a_{1,0}\;a_{2,1} + a_{0,1}\;a_{3,0} + 3\;a_{1,1}\;a_{2,0}$ & $3\;a_{1,0}^2\;a_{1,1} + 3\;a_{1,0}\;a_{0,1}\;a_{2,0}$ & $a_{1,0}^3\;a_{0,1}$ &\centering---&\\\\[-5pt]
$Q_{2,2}$ & $2\;a_{1,0}\;a_{1,2} + 2\;a_{0,1}\;a_{2,1} + a_{2,0}\;a_{0,2} + 2\;a_{1,1}^2$ & $a_{1,0}^2\;a_{0,2} + 4\;a_{1,0}\;a_{0,1}\;a_{1,1} +a_{0,1}^2\;a_{2,0}$ & $a_{1,0}^2\;a_{0,1}^2$ &\centering---&\\\\[-5pt]
\hline\\[-5pt]
$Q_{5,0}$ & $10 \;a_{2,0} \;a_{3,0} + 5 \;a_{1,0} \;a_{4,0}$ & $15 \;a_{1,0} \;a_{2,0}^2 + 10 \;a_{1,0}^2 \;a_{3,0}$ & $10 \;a_{1,0}^3 \;a_{2,0}$ & $a_{1,0}^5$ &\\\\[-5pt]
$Q_{4,1}$ & $6 \;a_{2,0} \;a_{2,1} + 4 \;a_{1,1} \;a_{3,0} + 4 \;a_{1,0} \;a_{3,1} + a_{0,1} \;a_{4,0}$ & $12 \;a_{1,0} \;a_{1,1} \;a_{2,0} + 3 \;a_{0,1} \;a_{2,0}^2 + 6 \;a_{1,0}^2 \;a_{2,1} + 4 \;a_{0,1} \;a_{1,0} \;a_{3,0}$ & $4 \;a_{1,0}^3 \;a_{1,1} + 6 \;a_{0,1} \;a_{1,0}^2 \;a_{2,0}$ & $a_{0,1} \;a_{1,0}^4$  &\\\\[-5pt]
$Q_{3,2}$ & $3 \;a_{1,2} \;a_{2,0} + 6 \;a_{1,1} \;a_{2,1} + 3 \;a_{1,0} \;a_{2,2} + a_{0,2} \;a_{3,0} + 2 \;a_{0,1} \;a_{3,1}$ & $6 \;a_{1,0} \;a_{1,1}^2 + 3 \;a_{1,0}^2 \;a_{1,2} + 3 \;a_{0,2} \;a_{1,0} \;a_{2,0} + 6 \;a_{0,1} \;a_{1,1} \;a_{2,0} + 6 \;a_{0,1} \;a_{1,0} \;a_{2,1} + a_{0,1}^2 \;a_{3,0}$ & $a_{0,2} \;a_{1,0}^3 + 6 \;a_{0,1} \;a_{1,0}^2 \;a_{1,1} + 3 \;a_{0,1}^2 \;a_{1,0} \;a_{2,0}$ & $a_{0,1}^2 \;a_{1,0}^3$ &\\\\[-7pt]
\hline
\end{tabular}
\caption{Coefficients $c_{n-p,n}^{k}$ (with $k$ referring to column $W_k$) as defined in \Eq{eq:ck}, for orders $2 \leq p\leq 5$. Since the formal expression of $Q_{i,j}$ is the same as $Q_{j,i}$ (but $Q_{i,j}\neq Q_{j,i}$), interchanging all indexes $(i,j)$ appearing in the expressions of $Q_{p-n,n}$ gives the expression of $Q_{n,p-n}$, so it is enough to provide $c_{n-p,n}^{k}$ for $n\leq \lfloor p/2\rfloor$.}
\end{table*}

	The main steps of the method are as follows. For a given position of the source centre at affix $\zeta_0=\xi_0+i\eta_0 \in\mathbb{C}$ \citep{Witt1990} in the source plane, I first expand the image position $z \in\mathbb{C}$ around the exact position of the image $z_0$ of the source in the lens plane. I then use this expansion to perform the integration of the area of the image through the  Green-Riemann formula, from which I obtain the magnification as a series of powers of $\rho^{2}$. 
	
	While the method itself seems fairly clear, in practice it is not straightforward to obtain simple expressions of the coefficients of the expansion, which should ideally be easy to calculate and fast to compute numerically, as it is of main interest in this work. I find that by using a combination of a Taylor expansion with respect to the two coordinates in the source plane $(\xi,\eta)$ and using properties of complex numbers provide an elegant and powerful way to express the expansion, as I show below. 
	
	Let us consider a microlensing system composed of $L$ components with mass ratio $q_l=M_l/M$ with respect to the total lens mass $M$, and located at positions $s_l\in\mathbb{C}$ in the lens plane. The lens equation then reads\footnote{A shear can be added in the equation, with minor changes in the expansion as explained in footnote \ref{note:shear}.}
\begin{equation}
\label{eq:lenseq}
	\zeta = z - \sum\limits_{l=1}^{L}\frac{q_l}{(\overline{z-s_l})} = z -\overline{W_1}\,,
\end{equation}
where I have introduced the $W_k$ factors, $k\geq 1$, as 
\begin{equation}
\label{eq:}
	W_k  \equiv (-1)^{(k-1)}(k-1)!\sum\limits_{l=1}^{L}\frac{q_l}{\left(z-s_l\right)^k} \,.
\end{equation}
For a point-source $\zeta$, the lens equation provides several images $j$ located at $z_j$ \citep[3 or 5 for binary lenses, $n+1$ to $5n-5$ for $n$ lens components,][]{Rhie2003}, but for clarity I have omitted any explicit reference to $j$ in $z$, $W_k$ and other quantities introduced later. The signed point-source magnification $\mu_0$ of an image is the inverse of the determinant of the Jacobi matrix of transformation $(z,\overline{z})\mapsto(\zeta,\overline{\zeta})$ \citep{Witt1990,DanekHeyrovsky2015}, given by $J\equiv\det\;\partial(\zeta,\overline{\zeta})/\partial (z,\overline{z})$, or
\begin{equation}
\label{eq:J}
	J = \frac{\partial\zeta}{\partial z}\frac{\partial\overline{\zeta}}{\partial \overline{z}}-\frac{\partial\zeta}{\partial\overline{z}}\frac{\partial\overline{\zeta}}{\partial z} = 1-\left|\frac{\partial\zeta}{\partial\overline{z}}\right|^2 = 1-\left|W_2\right|^2 \,.
\end{equation}
Hence the point-source magnification reads 
\begin{equation}
\label{eq:mu0}
	\mu_0 = \frac{1}{1-\left|W_2\right|^2} \,,
\end{equation}
the sign of which (in fact, the sign of $J$) gives the parity of the image. The critical curves correspond to infinite values of $\mu_0$, or $W_2=-e^{-i\phi}$, with $\phi$ a phase parameter ranging from $0$ to $2\pi$ \citep[][see Appendix \ref{sec:app} for two interpretations of $\phi$]{Witt1990}.

	I proceed now with the first step, the expansion of $z$ around $z_0$, the exact image of the source centre $\zeta_0$. At order $p\geq 1$, the Taylor expansion of $z$ as a function of coordinates $(\xi,\eta)\in\mathbb{R}^2$ in the source plane is
\begin{equation}
\label{eq:zdev}
	 \displaystyle z\left(\xi,\eta\right) = z\left(\xi_0+\delta\xi,\eta_0+\delta\eta\right) \simeq z_0 +\sum\limits_{p \geq 1} \frac{1}{p!}\left[\frac{\partial z}{\partial\xi}\;\delta\xi+\frac{\partial z}{\partial\eta}\;\delta\eta\right]^{[p]}\,,
\end{equation}
in which $[p]$ refers to the symbolic binomial expansion of the terms inside the brackets\footnote{For example, $[p]=2$ gives $\displaystyle \frac{\partial^2 z}{\partial\xi^2}\;\delta\xi^2+2\,\frac{\partial^2 z}{\partial\xi\partial\eta}\;\delta\xi\delta\eta+\frac{\partial^2 z}{\partial\eta^2}\;\delta\eta^2$.}, and where the derivatives are evaluated at $(\xi_0,\eta_0)$. In the following, I will use a more compact notation of the derivatives, 
\begin{equation}
\label{eq:amn}
	a_{p-n,n} = \frac{\partial^p z}{\partial\xi^{p-n}\partial\eta^n}\,,
\end{equation}
for $0\leq n\leq p$. Until this point, I have used the linearity of the complex notation as a convenient way to write the Taylor expansion of $z=x+iy$. A naive approach would be to differentiate $x$ and $y$ with respect to $\xi$ and $\eta$, but this requires to separate real and imaginary parts of the lens equation \Eq{eq:lenseq}, which results in cumbersome, numerically time-consuming expressions of the derivatives, and furthermore depends on the detailed form of the adopted lens equation. In the following, I will therefore use the complex formalism and exploit the property that $W_k$ is a single function of $z$, so that\footnote{\label{note:shear}The method can be extended to any expression $W_k$ depending on $z$ only, as for example adding a shear $\gamma$. Then, $\zeta=z+\gamma\overline{z}-\overline{W_1}$, $W_1^\prime = -\gamma z + W_1$, $W_2^\prime = -\gamma + W_2$ and for $k\geq 3$, $W_k^\prime = W_k$.}
\begin{equation}
\label{eq:DWk}
	\frac{\textrm{d} W_k}{\textrm{d} z} = W_{k+1}\,.
\end{equation}	
I  use this property to compute the derivatives \Eq{eq:amn}, starting with $p=1$ for which I provide the explicit expressions, and use $p=2$ to explain the general method for any $p$.

	From the lens equation, one has 
\begin{equation}
\label{eq:zbase}
	\frac{\partial\zeta}{\partial\alpha} = \frac{\partial z}{\partial\alpha}-\overline{W_2}\frac{\partial \overline{z}}{\partial\alpha}\,,
\end{equation}
where the differentiation is made with respect to $\alpha\in\{\xi,\eta\}$. To eliminating derivatives of $\overline{z}$, I conjugate \Eq{eq:zbase}, isolate $\partial\overline{z}/\partial\alpha$ and introduce it back into \Eq{eq:zbase}, which leads to
\begin{equation}
\label{eq:tmp1}
	\frac{\partial z}{\partial\alpha}\left(1-W_2\overline{W_2}\right) = \frac{\partial\zeta}{\partial\alpha} + \overline{W_2}\frac{\partial \overline{\zeta}}{\partial\alpha}\,.
\end{equation}
Considering that $1-W_2\overline{W_2}=1/\mu_0$ and that
\begin{align}
\label{eq:Dpzeta}
	\displaystyle \frac{\partial\zeta}{\partial\xi}=1\,,\quad
	\frac{\partial\zeta}{\partial\eta}=i\,,\quad
	\forall p>1,\; \frac{\partial^{p}\zeta}{\partial\xi^{p-n}\partial\eta^n}=0\,,
\end{align}
I obtain the two derivatives $a_{1,0}$ and $a_{0,1}$ ($p=1$) by successively choosing $\alpha=\xi$ and $\eta$,
\begin{align}
\label{eq:a10a01}
	a_{1,0} =&  \mu_0\left(1+\overline{W_2}\right)\,, \\\nonumber
	a_{0,1} =&  i\mu_0\left(1-\overline{W_2}\right)\,.
\end{align}

\begin{figure*}
\begin{center}
\includegraphics[width=\textwidth]{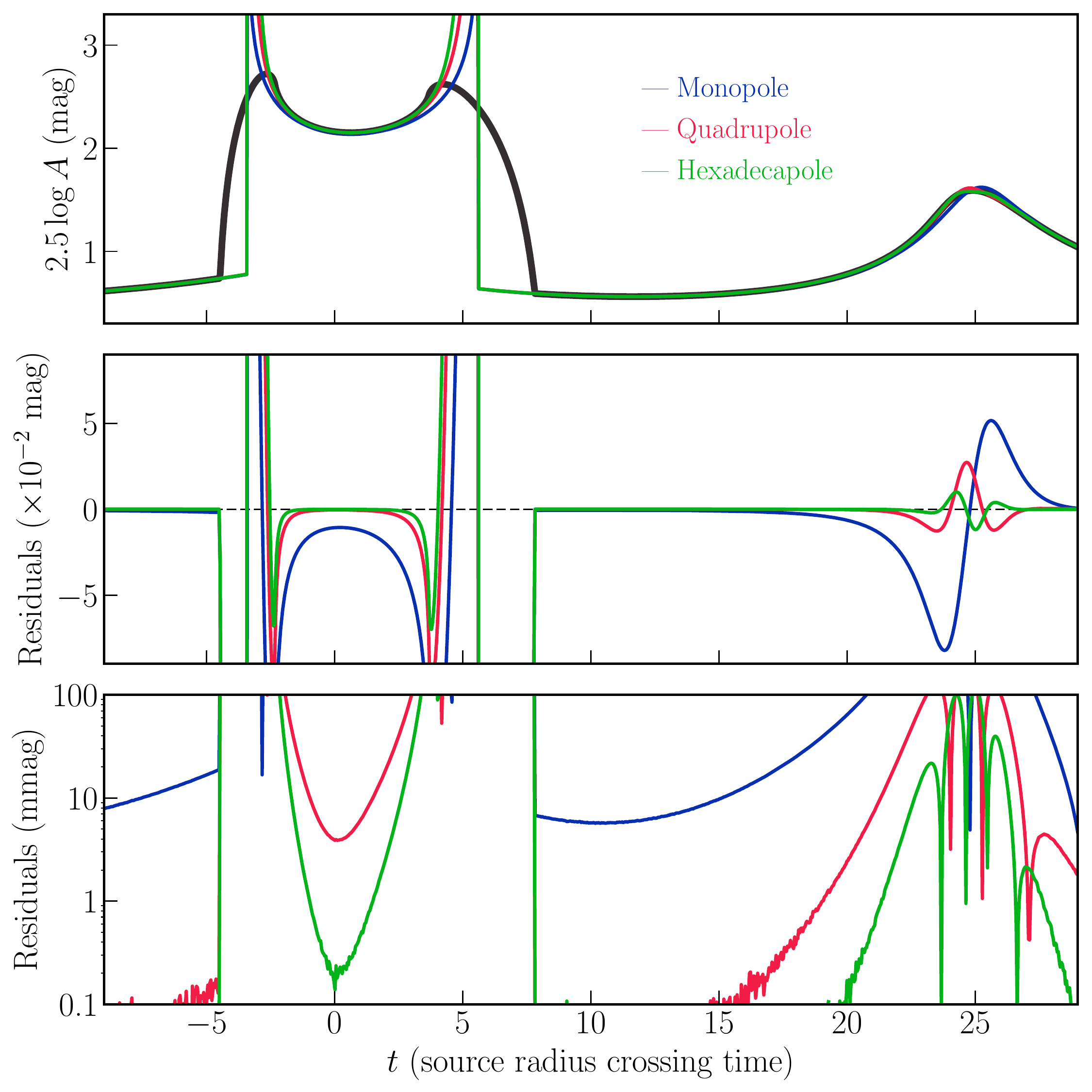}
\caption{Upper panel: simulated microlensing light curve (magnitude of total magnification $A$ as a function of time $t$ in units of source radius crossing time, for $s=1.7$, $q=0.2$ and for a uniformly bright source) displaying two clear caustic crossings ($t\sim -4$ and $t\sim 8$) and a cusp approach ($t\sim 25$). The exact finite-source magnification is the bold, dark grey curve, while the blue, red and green curves are respectively the monopole, quadrupole and hexadecapole approximations. 
Middle panel: residual magnitudes of the different finite-source approximations with respect to the exact magnification. Lower panel: absolute value of the residuals in logarithmic scale, for reference to future space-based microlensing missions which are expected to reach mmag precision light curve photometry. The small wiggles below the fraction of mmag come from errors in computing the exact magnification rather than from the multipole approximations.}
\label{fig:compar}
\end{center}
\end{figure*}

	For $p=2$, I derive \Eq{eq:zbase} a second time with respect to variables $(\alpha,\beta)\in\{\xi,\eta\}\times\{\xi,\eta\}$, which leads to three different combinations $a_{2-n,n}$ ($n=0$, 1 and 2) satifying
\begin{equation}
\label{eq:z2}
	\frac{\partial^2\zeta}{\partial\alpha\partial\beta} = \frac{\partial^2 z}{\partial\alpha\partial\beta}-\frac{\partial \overline{W_2}}{\partial\beta}\frac{\partial \overline{z}}{\partial\alpha}-\overline{W_2}\frac{\partial^2 \overline{z}}{\partial\alpha\partial\beta}=0\,.
\end{equation}
Using the rule \Eq{eq:DWk}, the derivative of $\overline{W_2}$ can be expressed as
\begin{equation}
	\frac{\partial \overline{W_2}}{\partial\beta} = \frac{d \overline{W_2}}{d \overline{z}} \frac{\partial \overline{z}}{\partial\beta} = \overline{W_3} \frac{\partial \overline{z}}{\partial\beta}\,,
\end{equation}
so that \Eq{eq:z2} involves derivatives of $z$ and $\overline{z}$, and constants depending on $z_0$ only. With the same procedure as for $p=1$ (conjugating and replacing), I get
\begin{align}
\label{ }
	a_{2,0} =&   \mu_0\;\left(\overline{W_3\,a_{1,0}^2}+\overline{W_2}\,W_3\,a_{1,0}^2\right)\,,\\\nonumber
	a_{1,1} =&   \mu_0\;\left(\overline{W_3\,a_{1,0}\;a_{0,1}}+\overline{W_2}\,W_3\,a_{1,0}\;a_{0,1}\right)\,,\\\nonumber
	a_{0,2} =&   \mu_0\;\left(\overline{W_3\,a_{0,1}^2}+\overline{W_2}\,W_3\,a_{0,1}^2\right)\,.
\end{align}
More generally for $p\geq 1$ and $0\leq n\leq p$, it appears that $a_{p-n,n}$ can be expressed as
\begin{equation}
\label{eq:aQ}
	a_{p-n,n} = \mu_0\;\left(\overline{Q_{p-n,n}} + \overline{W_2}\,Q_{p-n,n}\right)\,,
\end{equation}
where the $Q_{p-n,n}$ coefficients can be iteratively calculated using the prescriptions given above. Factoring the $W_k$, I define 
\begin{equation}
\label{eq:ck}
	Q_{p-n,n} = \sum_{k=3}^{p+1} c_{n-p,n}^{k}\,W_k\,, 
\end{equation}
where the coefficients $c_{n-p,n}^{k}$ are given in \Tab{tab:Q} up to order $p=5$ (we shall see later that expanding $z$ up to this order provides the hexadecapole term of the finite-source magnification). For $p\geq 3$, I find that $Q_{p-n,n}$ can be obtained with the following algorithm: let us introduce $p$ variables $(\alpha_p,\dots, \alpha_1)\in\{\xi,\eta\}\times\dots\times\{\xi,\eta\}$. For $3\leq k\leq p+1$, the general expression of $c_{n-p,n}^{k}$ is obtained from
\begin{equation}
\label{eq:algoRkp}
	R_p^k = \frac{\partial R^k_{p-1}}{\partial \alpha_p} + \left(R_{p-1}^{k-1} + \delta(k,3) \frac{\partial^{p-1} z}{\partial\alpha_{p-1}\dots\partial\alpha_1}\right) \frac{\partial z}{\partial\alpha_p}\,,
\end{equation}
with $\delta(3,k)=1$ only if $k= 3$, and in which $n$ of the $\alpha_i$ variables are chosen to be $\eta$, and the remaining $p-n$ others $\xi$. 

	The second step of the method consists of calculating the area of the image using the previous expansion of $z$, with the provisional assumption that the source is uniformly bright. Let $\rho$ be the source radius in Einstein units. For $\rho\ll 1$, the (circular) contour of the source can be parametrized as $\zeta=\zeta_0+\rho\cos\theta +i\,\rho\sin\theta$, i.e. by choosing 
$\delta\xi=\rho\cos\theta$ and $\delta\eta=\rho\sin\theta$ in the expansion of $z$ written in \Eq{eq:zdev}. As a matter of fact, $z$ is now a function of $(\rho,\theta)$, and to avoid confusions I introduce
\begin{equation}
\label{eq:Zre}
	Z(\rho,\theta) \equiv z\big(\xi_0+\rho\cos\theta,\eta_0+\rho\sin\theta\big)-z_0\,.
\end{equation}
This expression hence involves powers of $\rho$ as well as powers of $\cos\theta$ and $\sin\theta$. For a given source size $\rho$, the (signed) area $S$ of the image can be performed through the Green-Riemann (or Stokes) formula,
\begin{equation}
\label{eq:S}
	S = \Im\left[\;\oint_\mathcal{C} \frac{\overline{Z}\,\textrm{d}Z}{2} \;\right] = \frac{1}{2}\Im\left[\;\int_0^{2\pi} \overline{Z}\,\frac{\partial Z}{\partial\theta} \; \textrm{d}\theta\;\right] \,,
\end{equation}
since for $Z=X+iY$ one has $-Y\textrm{d}X+X\textrm{d}Y=\Im[\overline{Z}\,\textrm{d}Z]$. As the expression of $Z$ obtained in \Eq{eq:Zre} is analytical, it is also the case for $\partial Z/\partial\theta$. The integral \Eq{eq:S} involves terms like $(\cos^a\theta\sin^b\theta)$, and the integration is easily handled formally with a software like \textsc{mathematica}. It is interesting to remark that all terms with odd values of $a$ or $b$ cancel out, so after the integration, only terms with even powers of $\rho$ remain. Furthermore, it can be found by inspecting $\overline{Z}$ and $\partial Z/\partial\theta$ that pursuing the expansion to order $p$ adds terms factors of $\rho^{m}$, where $m\geq p+1$. In other words, the expansion of $Z$ is complete in $\rho^{m}$ for $p=m-1$. 

	Expanding $S$ in the form 
\begin{equation}
\label{eq:defS}
	S(\rho) = \sum\limits_{p\geq 1}\frac{\pi\rho^{p+1}}{(p-1)!}\,\mu_{p-1}
\end{equation}
yields, up to order $p=5$, the following non-vanishing terms
\begin{align}
\label{eq:mu0}
\mu_0 = & \;\Im\Big[\overline{a_{1,0}}\;a_{0,1}\Big]\,,\\
\label{eq:mu2}
\mu_2 = &\;\displaystyle\frac{1}{4}\; \Im\Big[\overline{a_{1,0}}\;(a_{2,1}+a_{0,3})+2\;(a_{1,1}\;\overline{a_{2,0}}+\overline{a_{1,1}}\;a_{0,2})\\\nonumber
	&\displaystyle+\;a_{0,1}\;(\overline{a_{1,2}}+\overline{a_{3,0}})\Big]\,,\\
\label{eq:mu4}
\mu_4 =&\;\displaystyle\frac{1}{8}\Im\Big[
	\overline{a_{1,0}}\;(a_{4,1}+2\;a_{2,3}+a_{0,5})+4\;\overline{a_{2,0}}\;(a_{1,3}+a_{3,1})\\\nonumber
	&\displaystyle+4\;a_{1,1}\;\overline{a_{4,0}}+6\;(a_{0,3}\;\overline{a_{1,2}}+a_{1,2}\;\overline{a_{2,1}}+a_{2,1}\;\overline{a_{3,0}})\\\nonumber	 
	&\displaystyle+2\;a_{0,3}\;\overline{a_{3,0}}+4\;\overline{a_{1,1}}\;a_{0,4}+4\;a_{0,2}\;(\overline{a_{1,3}}+\overline{a_{3,1}})\\\nonumber
	&\displaystyle+a_{0,1}\;(\overline{a_{1,4}}+2\;\overline{a_{3,2}}+\overline{a_{5,0}}) \Big]\,.
\end{align}
The first term $\mu_0$ is indeed the point-source magnification defined in \Eq{eq:mu0}, since from \Eq{eq:a10a01} one has $\overline{a_{1,0}}\;a_{0,1}=i\mu_0^2(1-|W_2|^2-\overline{W_2}+W_2)$ of which imaginary part is $\mu_0^2(1-\left|W_2\right|^2)=\mu_0$.  Let us now consider a limb-darkened source with brightness profile
\begin{equation}
\label{ }
	I(r) = 1-\Gamma  \left(1-\frac{3}{2} \sqrt{1-\frac{r^2}{\rho ^2}}\right)\,,\quad 0\leq r\leq\rho\,,
\end{equation}
where for all $\Gamma$, the surface integral of $I(r)$ over the source face $\mathcal{S}$ always equals $\pi\rho^2$,
\begin{equation}
\label{ }
	\displaystyle\iint_\mathcal{S}I(r)\,\textrm{d}s = 
	\int_{0}^{2\pi}\textrm{d}\phi\int_{0}^{\rho}I(r)r\textrm{d}r = \pi\rho^2 \,.
\end{equation} 
The uniformly bright source has $\Gamma=0$. The signed magnification is now given by the ratio 
\begin{equation}
\label{ }
	\displaystyle\mu \equiv \frac{\int_{0}^{\rho} I(r)\,\textrm{d}S}{\iint_\mathcal{S} I(r)\,\textrm{d}s} =
	\frac{1}{\rho^2}\sum\limits_{p\geq 1}\frac{(p+1)\;\mu_{p-1}}{(p-1)!}\int_{0}^{\rho} I(r)\,r^p\,\textrm{d}r\,,
\end{equation}
where the enumerator of the first integral is already integrated over the angle, and in which I have changed variable $\textrm{d}S=\frac{\textrm{d}S}{\textrm{d}r}\,\textrm{d}r$ according to \Eq{eq:defS}. The integral can be performed analytically for any value of $p$, and yields
\begin{equation}
\label{eq:devmu}
	\mu = \mu_0 + \frac{\mu_2}{2!}\left(1-\frac{1}{5}\Gamma\right)\,\rho^2+\frac{\mu_4}{4!}\left(1-\frac{11}{35}\Gamma\right)\,\rho^4 + \mathcal{O}\left(\rho^6\right)\,.
\end{equation}
As expected, the monopole of the finite-source expansion ($p=1$) is the point-source magnification of the source, and there is no dipole ($p=2$). The quadrupole is obtained for $p=3$, and the hexadecapole for $p=5$.

	Finally, the total magnification of the source $A(\zeta_0)$ is the sum of the absolute values of the individual magnification factors $\mu(z_j)$ of each of the images $j$. If $\epsilon_j$ denotes the parity of image $j$ ($\epsilon_j= 1$ if $\mu>0$, $-1$ otherwise), one has
\begin{equation}
\label{eq:mufin}
	A = \sum\limits_j\big|\mu(z_j)\big| = \sum_j \epsilon_j\,\mu(z_j)  \,,
\end{equation}
so that after factorizing $\rho^2$ and $\rho^4$ amongst the different images, the total magnification $A$ has the same form\footnote{This is the same expansion as \cite{Gould2008} but with a different choice of numerical factors.} as \Eq{eq:devmu} with $A_{q}$ instead of $\mu_{q}$, where $A_q$ is a combination of $\epsilon_j \mu_{q}$.

\section{Application} \label{sec:appli}

	An example of light curves obtained with the various finite-source approximations are displayed in \Fig{fig:compar}. In this example, the lens is a binary with parameters $s=1.7$ (separation in Einstein units) and $q=0.2$ (lens mass ratio), and the source radius is $\rho=0.01$ in Einstein units (slightly larger than typical source sizes to better see the differences). The source crosses two caustics (entry at $t\sim -4$, exit at $t\sim 8$) and later approaches a cusp ($t\sim 25$). The exact finite-source magnification is displayed as the bold, dark grey curve, while the blue, red and green curves are respectively the monopole, quadrupole and hexadecapole approximations. The middle panel shows a zoom on the residuals (approximated minus exact magnifications) in linear scale, while the lower panel shows the absolute value of the residuals in logarithmic scale. The latter panel can be used to compare the precision of the approximations to the precision of the photometry expected from future space-based missions such as \textit{Wide-Field Infrared Survey Telescope}, which is expected to be of order of a mmag. 
	
	It appears that the quadrupole expansion already provides a much better approximation than the monopole (point-source magnification), in particular, near cusp approaches. The hexadecapole appears essentially as an approximation that allows the source to approach the caustics slightly closer than the quadrupole before the approximation breaks down \citep[see also the discussion in][]{Gould2008}. 
	
	Since one of the drivers of this work was to improve the numerical efficiency in calculating the quadrupole and hexadecapole approximations, I have tested a non-fully optimized python routine to estimate the potential gain for a binary lens. I find that the quadrupole (respectively hexadecapole) is about $\times\,2$ (respectively $\times\,5$) slower than point source. The quadrupole and hexadecapole implementations presented here are respectively $\times\,6$ and $\times\,4$ faster than the implementation of \cite{Gould2008}.
It is likely that going to higher orders in the finite-source approximation will not help, though, not only because the gain in precision will be limited, but also because the additional calculations (such as $a_{p-n,n}$) grow substantially with increasing order $p$. It is also clear that the gain in time will increase with the number of lens components since solving the lens equation will take more time, while the additional calculations do not, as they depend only on the set of input complex numbers $W_k$. 

	Additionally, a further advantage of using the exact multipole expansion rather than a numerical estimation resides in the fact that the discontinuities of the magnification happen at the same positions in the light curve than with point-source, so there are no spurious wiggles of the magnification close to the caustics.

	An open source code with a first implementation of the equations presented here is available for download in my GitHub repository\footnote{\url{https://github.com/ArnaudCassan/microlensing/}}. The file \verb"multipoles.py" includes two functions that return the quadrupole and hexadecapole approximations of the finite-source magnification: \verb"quadrupole(Wk,rho,Gamma)" and \verb"hexadecapole(Wk,rho,Gamma)", whose arguments are $W_k$ (a 2D complex array in which one of the dimension refers to the individual images), $\rho$ the source radius in Einstein units and $\Gamma$ the source linear limb-darkening coefficient. In practice (as shown in the \verb"example()" function provided), for each position of the source, one needs to compute the images of the source centre, discard the virtual ones, and evaluate the corresponding $W_k$ (up to $k=4$ for the quadrupole and $k=6$ for the hexadecapole) that are inputs of the two functions.  It is likely that these functions will gain in speed from a re-writing in \textsc{cython} or \textsc{c++} with a more efficient use of complex number calculations. For information, further orders of $a_{p-n,n}$ for $p\geq 6$ can be displayed using function \verb"Q(p)" in file \verb"Rkp.py", as an implementation of \Eq{eq:algoRkp}.
	
\section{Conclusion}

	I have presented a method that allows to compute the quadrupole and hexadecapole approximations with more efficiency than previously available codes (respectively about six times and four times faster). It appears that the quadrupole approximation already provides a much more precise approximation than point-source, for a computing time only about two times slower. This advocates for using the quadrupole in place of point-source in most part of the light curve, except at baseline. The hexadecapole seems well suited to make the link between exact finite-source and quadrupole in limited regions of the light curves close to sharp magnification peaks. Open source codes of the algorithms presented here are available for download, and I welcome numerical optimization updates.
 
\section*{Acknowledgements}

AC acknowledges financial support from Université Pierre et Marie Curie under grants Émergence@Sorbonne Universités 2016 and Émergence-UPMC 2012. I would like to thank the anonymous referee for useful remarks on the text.

\appendix
\section{Two interpretations of Witt's $\phi$}\label{sec:app}

	In his original article, \cite{Witt1990} introduced a phase parameter $\phi\in[0,2\pi]$ to compute the critical curves\footnote{The two $-$ signs reflect the different conventions between the formalisms.}, 
\begin{equation}
\label{eq:phiW2}
	W_2=-e^{-i\phi}\,.
\end{equation}
In this appendix I propose two interpretations of $\phi$. 

	The first one is geometrical. Starting from the lens equation \Eq{eq:lenseq}, one can write
\begin{equation}
\label{eq:dezp}
	\frac{d\zeta}{d\phi} = \frac{dz}{d\phi}-\overline{W_2}\frac{d\overline{z}}{d\phi}\,,
\end{equation}
which, after conjugating the whole expression and replacing $d\overline{z}/d\phi$ in \Eq{eq:dezp}, yields
\begin{equation}
\label{}
	\frac{d\zeta}{d\phi} = e^{i\phi}\frac{d\overline{\zeta}}{d\phi}\,.
\end{equation}
Since $d\zeta/d\phi$ defines in the source plane a vector $T$ tangent to the caustic curve parametrized by $\phi$, the former equation means that $\phi$ is the oriented angle between the symmetrical vector of $T$ with respect to the horizontal axis and $T$. In other words, $\phi/2$ modulo $\pi$ is the geometric angle between the horizontal axis and the tangent to the caustic.

	A second interpretation of $\phi$ is related to curves in the lens plane obtained with $\phi={\rm cst}$. From the definition of the Jacobian $J$ in \Eq{eq:J} and since $J<1$, one has
\begin{equation}
\label{ }
	J = 1-W_2\overline{W_2}=1 - \left(-e^{-i\phi}\sqrt{1-J}\right)\left(-e^{i\phi}\sqrt{1-J}\right)\,,
\end{equation}
from which we write 
\begin{equation}
\label{eq:W2phi}
	W_2=-e^{-i\phi}\sqrt{1-J}\,.
\end{equation}
Solving this equation for a given $\phi\in[0,2\pi[$ and $J\in]-\infty, 1[$ leads to four possible solutions $z$ in the lens plane. Varying $\phi$ for a given $J$ draws iso-magnification curves (as four distinct branches that connect), that I will refer to as $J$-curves. Let us study the curves obtained for $\phi$ constant. Differentiating both sides of equation \Eq{eq:W2phi} with respect to $\phi$ and $J$ yields
\begin{equation}
\label{eq:twopartial}
	W_3\frac{\partial z}{\partial\phi} = -iW_2\,, \quad W_3\frac{\partial z}{\partial J} = -\frac{W_2}{2(1-J)}\,,
\end{equation}
and
\begin{equation}
\label{ }
	\frac{\partial z}{\partial\phi} = i2\left(1-J\right)\,\frac{\partial z}{\partial J}\,.
\end{equation}
Since $\partial z/\partial J$ defines in the lens plane a vector $T_J$ tangent to the  $J$-curves, $\partial z/\partial\phi$ hence defines a vector $T_\phi$ perpendicular to those, as long as $T_\phi\neq 0$ (or equivalently $T_J\neq 0$). It means that $\phi={\rm cst}$ lines can be interpreted as field lines (perpendicular to the $J$-curves) parametrized by $J$ and crossing the critical lines at $J=0$. I will call them $\phi$-lines. 

\begin{figure}
\begin{center}
\includegraphics[width=\columnwidth]{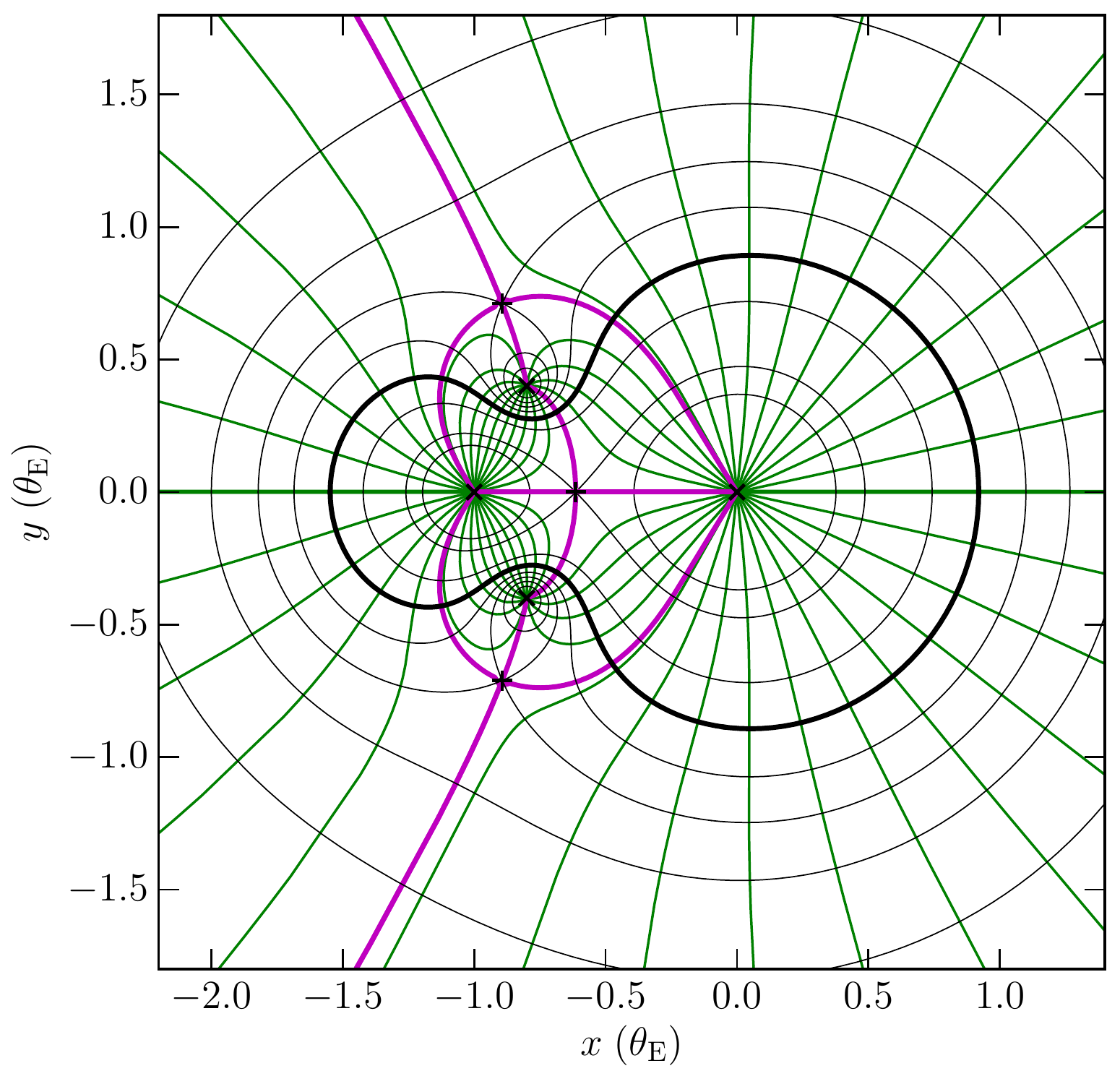}
\caption{Binary lens with the less massive component on the left hand side ($z_2=-s$, with $z=x+iy$) and the  more massive body at the centre of the coordinate system ($z_1=0$), the positions of which are marked as black $\times$ on the horizontal axis. The relative mass ratio of the system is $q=\mu_2/\mu_1<1$. The thick black curve is the critical line ($J=0$), and the thin black lines are the $J$-curves ($J={\rm cst}$). The $\phi$-lines ($\phi={\rm cst}$) are plotted as thin green lines. They are orthogonal to the $J$-curves except at saddle points, which are marked as the two off-axis black $+$. The two off-axis black $\times$ mark the two extrema (maxima) of $J(x,y)$. The thick magenta lines mark the boundaries between the different kinds of field lines (finite or semi-infinite, starting at lens positions or at extrema of $J$). There are six of them in the binary lens case, which all (necessarily) pass through a saddle point.}
\label{fig:isofield}
\end{center}
\end{figure}

	An example is given in  \Fig{fig:isofield} for a binary lens with separation $s=1$ and mass ratio $q=0.25$. The $\phi$-lines are the green, radial lines crossing the (thin black) $J$-curves orthogonally (except at saddle points, marked as black $+$, see below). The thick black line is the critical curve. $\phi$-lines can be of finite length or semi-infinite. For $J\rightarrow-\infty$, the only possibility is that at least one of the $z\rightarrow s_l$, so that all $\phi$-lines start at one of the lens component positions. For $J\rightarrow1$, one has $W_2\rightarrow0$, which results in two possibilities for $z$: either $z\rightarrow\infty$ and the line goes to infinity, or $z$ converges to a finite value, which geometrically  necessarily correspond to an extrema of  mapping $J(z)$. Expanding \Eq{eq:W2phi} and setting $\sqrt{1-J}\rightarrow0$, one finds\begin{equation}
\label{eq:saddle}
\sum_{l=1}^L q_l \prod_{k\neq l} (z-z_k)^2 = 0\,.
\end{equation}
Therefore for $L$ lens components, there are $2(L-1)$ extrema of $J(z)$. In the binary lens case, with the convention that the origin of the coordinate system is at the position of the more massive body $z_1=0$ and that the less massive body is located at $z_2=-s$, with $q=\mu_2/\mu_1<1$ their relative mass ratio, there are two extrema at
\begin{equation}
\label{ }
	z_\pm=-\frac{s}{1+q}\left(1\pm i\sqrt{q}\right)\,,
\end{equation} 
which are marked as off-axis $\times$ in \Fig{fig:isofield}.

	Finally, the boundary values of $\phi$ which separate regions of different kinds of $\phi$-lines (finite or semi-infinite, starting at one or another lens position) correspond to lines that necessarily cross the $J$-curves at saddle points as stated before, where$W_3=0$ \citep{DanekHeyrovsky2015} i.e.  
\begin{equation}
\label{ }
\sum_{l=1}^L q_l \prod_{k\neq l} (z-z_k)^3 = 0\,.
\end{equation}
The boundary $\phi$-lines may start and/or end at one of the lens position, at a maxima of $J(z)$ or at infinity, but they all necessarily pass through a saddle point. Once the saddle points are found through \Eq{eq:saddle}, the corresponding values of $\phi$ are obtained from \Eq{eq:phiW2} and the corresponding boundary $\phi$-lines can be drawn. In the binary lens case, it appears that there exists six such boundary $\phi$-lines, displayed in \Fig{fig:isofield} as thick magenta lines.




\bibliographystyle{mnras}
\bibliography{bibli} 

\begin{thebibliography}{}
\makeatletter
\relax
\def\mn@urlcharsother{\let\do\@makeother \do\$\do\&\do\#\do\^\do\_\do\%\do\~}
\def\mn@doi{\begingroup\mn@urlcharsother \@ifnextchar [ {\mn@doi@}
  {\mn@doi@[]}}
\def\mn@doi@[#1]#2{\def\@tempa{#1}\ifx\@tempa\@empty \href
  {http://dx.doi.org/#2} {doi:#2}\else \href {http://dx.doi.org/#2} {#1}\fi
  \endgroup}
\def\mn@eprint#1#2{\mn@eprint@#1:#2::\@nil}
\def\mn@eprint@arXiv#1{\href {http://arxiv.org/abs/#1} {{\tt arXiv:#1}}}
\def\mn@eprint@dblp#1{\href {http://dblp.uni-trier.de/rec/bibtex/#1.xml}
  {dblp:#1}}
\def\mn@eprint@#1:#2:#3:#4\@nil{\def\@tempa {#1}\def\@tempb {#2}\def\@tempc
  {#3}\ifx \@tempc \@empty \let \@tempc \@tempb \let \@tempb \@tempa \fi \ifx
  \@tempb \@empty \def\@tempb {arXiv}\fi \@ifundefined
  {mn@eprint@\@tempb}{\@tempb:\@tempc}{\expandafter \expandafter \csname
  mn@eprint@\@tempb\endcsname \expandafter{\@tempc}}}

\bibitem[\protect\citeauthoryear{{Albrow} et~al.,}{{Albrow}
  et~al.}{1999}]{Albrow1999}
{Albrow} M.~D.,  et~al., 1999, \apj, 522, 1022

\bibitem[\protect\citeauthoryear{{Bennett}}{{Bennett}}{2010}]{Bennett2010}
{Bennett} D.~P.,  2010, \mn@doi [\apj] {10.1088/0004-637X/716/2/1408}, \href
  {http://cdsads.u-strasbg.fr/abs/2010ApJ...716.1408B} {716, 1408}

\bibitem[\protect\citeauthoryear{{Bozza}}{{Bozza}}{2010}]{Bozza2010}
{Bozza} V.,  2010, \mn@doi [\mnras] {10.1111/j.1365-2966.2010.17265.x}, \href
  {http://adsabs.harvard.edu/abs/2010MNRAS.408.2188B} {408, 2188}

\bibitem[\protect\citeauthoryear{{Cassan}}{{Cassan}}{2008}]{Cassan2008}
{Cassan} A.,  2008, \aap, 491, 587

\bibitem[\protect\citeauthoryear{{Cassan}, {Horne}, {Kains}, {Tsapras}  \&
  {Browne}}{{Cassan} et~al.}{2010}]{Cassan2010}
{Cassan} A.,  {Horne} K.,  {Kains} N.,  {Tsapras} Y.,   {Browne} P.,  2010,
  \mn@doi [\aap] {10.1051/0004-6361/200913755}, \href
  {http://adsabs.harvard.edu/abs/2010A%26A...515A..52C} {515, A52}

\bibitem[\protect\citeauthoryear{{Cassan} et~al.,}{{Cassan}
  et~al.}{2012}]{Cassan2012}
{Cassan} A.,  et~al., 2012, \mn@doi [\nat] {10.1038/nature10684}, \href
  {http://adsabs.harvard.edu/abs/2012Natur.481..167C} {481, 167}

\bibitem[\protect\citeauthoryear{{Dan{\v e}k} \& {Heyrovsk{\'y}}}{{Dan{\v e}k}
  \& {Heyrovsk{\'y}}}{2015}]{DanekHeyrovsky2015}
{Dan{\v e}k} K.,  {Heyrovsk{\'y}} D.,  2015, \mn@doi [\apj]
  {10.1088/0004-637X/806/1/63}, \href
  {http://cdsads.u-strasbg.fr/abs/2015ApJ...806...63D} {806, 63}

\bibitem[\protect\citeauthoryear{{Dominik}}{{Dominik}}{2007}]{Dominik2007}
{Dominik} M.,  2007, \mnras, 377, 1679

\bibitem[\protect\citeauthoryear{{Dong} et~al.,}{{Dong}
  et~al.}{2006}]{Dong2006}
{Dong} S.,  et~al., 2006, \apj, 642, 842

\bibitem[\protect\citeauthoryear{{Gould}}{{Gould}}{2008}]{Gould2008}
{Gould} A.,  2008, \apj, 681, 1593

\bibitem[\protect\citeauthoryear{{Gould} \& {Gaucherel}}{{Gould} \&
  {Gaucherel}}{1997}]{GouldGaucherel1997}
{Gould} A.,  {Gaucherel} C.,  1997, \apj, \href
  {http://adsabs.harvard.edu/abs/1997ApJ...477..580G} {477, 580}

\bibitem[\protect\citeauthoryear{{Henderson} et~al.,}{{Henderson}
  et~al.}{2016}]{Henderson2016}
{Henderson} C.~B.,  et~al., 2016, \mn@doi [\pasp]
  {10.1088/1538-3873/128/970/124401}, \href
  {http://adsabs.harvard.edu/abs/2016PASP..128l4401H} {128, 124401}

\bibitem[\protect\citeauthoryear{{Kains}, {Browne}, {Horne}, {Hundertmark}  \&
  {Cassan}}{{Kains} et~al.}{2012}]{Kains2012}
{Kains} N.,  {Browne} P.,  {Horne} K.,  {Hundertmark} M.,   {Cassan} A.,  2012,
  \mn@doi [\mnras] {10.1111/j.1365-2966.2012.21813.x}, \href
  {http://adsabs.harvard.edu/abs/2012MNRAS.426.2228K} {426, 2228}

\bibitem[\protect\citeauthoryear{{Kubas} et~al.,}{{Kubas}
  et~al.}{2008}]{Jovi2008}
{Kubas} D.,  et~al., 2008, \mn@doi [\aap] {10.1051/0004-6361:20077449}, \href
  {http://adsabs.harvard.edu/abs/2008A%26A...483..317K} {483, 317}

\bibitem[\protect\citeauthoryear{{Liebig}, {D'Ago}, {Bozza}  \&
  {Dominik}}{{Liebig} et~al.}{2015}]{Liebig2015}
{Liebig} C.,  {D'Ago} G.,  {Bozza} V.,   {Dominik} M.,  2015, \mn@doi [\mnras]
  {10.1093/mnras/stv733}, \href
  {http://cdsads.u-strasbg.fr/abs/2015MNRAS.450.1565L} {450, 1565}

\bibitem[\protect\citeauthoryear{{Mao} \& {Paczynski}}{{Mao} \&
  {Paczynski}}{1991}]{MaoPaczynski1991}
{Mao} S.,  {Paczynski} B.,  1991, \apjl, 374, L37

\bibitem[\protect\citeauthoryear{{Pejcha} \& {Heyrovsk{\'y}}}{{Pejcha} \&
  {Heyrovsk{\'y}}}{2009}]{PejchaHeyrovsky2008}
{Pejcha} O.,  {Heyrovsk{\'y}} D.,  2009, \mn@doi [\apj]
  {10.1088/0004-637X/690/2/1772}, \href
  {http://adsabs.harvard.edu/abs/2009ApJ...690.1772P} {690, 1772}

\bibitem[\protect\citeauthoryear{{Penny}}{{Penny}}{2014}]{Penny2014}
{Penny} M.~T.,  2014, \mn@doi [\apj] {10.1088/0004-637X/790/2/142}, \href
  {http://cdsads.u-strasbg.fr/abs/2014ApJ...790..142P} {790, 142}

\bibitem[\protect\citeauthoryear{{Rhie}}{{Rhie}}{2003}]{Rhie2003}
{Rhie} S.~H.,  2003, preprint, \href
  {http://adsabs.harvard.edu/abs/2003astro.ph..5166R} {} (\mn@eprint {arXiv}
  {astro-ph/0305166})

\bibitem[\protect\citeauthoryear{{Skowron} \& {Gould}}{{Skowron} \&
  {Gould}}{2012}]{Skowron2012}
{Skowron} J.,  {Gould} A.,  2012, preprint, \href
  {http://cdsads.u-strasbg.fr/abs/2012arXiv1203.1034S} {} (\mn@eprint {arXiv}
  {1203.1034})

\bibitem[\protect\citeauthoryear{{Street} et~al.,}{{Street}
  et~al.}{2016}]{Street2016}
{Street} R.~A.,  et~al., 2016, \mn@doi [\apj] {10.3847/0004-637X/819/2/93},
  \href {http://adsabs.harvard.edu/abs/2016ApJ...819...93S} {819, 93}

\bibitem[\protect\citeauthoryear{{Udalski} et~al.,}{{Udalski}
  et~al.}{2015}]{Udalski2015}
{Udalski} A.,  et~al., 2015, \mn@doi [\apj] {10.1088/0004-637X/799/2/237},
  \href {http://cdsads.u-strasbg.fr/abs/2015ApJ...799..237U} {799, 237}

\bibitem[\protect\citeauthoryear{{Wambsganss}}{{Wambsganss}}{1997}]{Wambsganss1997}
{Wambsganss} J.,  1997, \mnras, 284, 172

\bibitem[\protect\citeauthoryear{Witt}{Witt}{1990}]{Witt1990}
Witt H.~J.,  1990, \aap, 236, 311

\makeatother
\end{thebibliography}




\bsp	
\label{lastpage}
\end{document}